\documentclass[11pt]{elsarticle}

\def	\be	{\begin{equation}}
\def	\ee	{\end{equation}}
\def	\bqt	{\begin{quote}}
\def	\eqt	{\end{quote}}
\def	\a	{\alpha}

\def	\mn 	{\mu \nu}
\def	\pl	{\partial}

\begin{document}

\title{On Coupling NEC-Violating Matter to Gravity}

\author{Saugata Chatterjee}

\author{Maulik Parikh}

\address{Department of Physics and Beyond: Center for Fundamental Concepts in Science \\
Arizona State University, Tempe, Arizona 85287, USA} 

\author{Jan Pieter van der Schaar}

\address{Delta Institute for Theoretical Physics, IOP, and GRAPPA\\ 
Universiteit van Amsterdam, Science Park 904, 1090 GL Amsterdam, Netherlands} 

\begin{abstract}
We show that effective theories of matter that classically violate the null energy condition cannot be minimally coupled to Einstein gravity without being inconsistent with both string theory and black hole thermodynamics. We argue however that they could still be either non-minimally coupled or coupled to higher-curvature theories of gravity.
\end{abstract}

\maketitle

\section{Introduction}
\label{introduction}

The null energy condition (NEC) requires that, at every point in spacetime,
\be
T_{\mn} v^\mu v^\nu \geq 0 \; ,	\label{TNEC}
\ee
for any light-like vector, $v^\mu$. The NEC is the weakest of the energy conditions; a violation of the NEC implies a violation of the weak, dominant, and strong energy conditions. The various energy conditions play a vital role in general relativity, where they are the main physical assumptions in the singularity theorems \cite{Penrose:1965} and in no-go theorems that prohibit the traversability of wormholes, the creation of laboratory universes \cite{Farhi:1987},  the building of time machines \cite{Hawking:1992}, and the possibility of bouncing cosmologies \cite{molinaparis,nobounce}. Perhaps most importantly, the energy conditions are needed in black hole thermodynamics \cite{Hawking:1976}: the NEC is used in proving that the area of a black hole event horizon, like entropy, always increases, while the dominant energy condition is used in proving the zeroth law, that the surface gravity of a black hole is uniform over the system at equilibrium, just like temperature.

Expressed in the form of (\ref{TNEC}), the NEC appears as a property of matter, since it is defined in terms of the matter energy-momentum tensor. All known forms of matter do obey the NEC classically, but one can wonder whether that has to be the case. Indeed, although NEC-violating theories often exhibit worrisome behavior, such as superluminal propagation \cite{Dubovsky:2005xd} or unbounded negative Hamiltonians \cite{alex}, they do not appear to be categorically ruled out by any principles of quantum field theory. In fact, it is easy to come up with a counter-example. Consider a free massless ghost. The Lagrangian density is
\be
{\cal L} = -\frac{1}{2} (\dot \varphi)^2 + \frac{1}{2} (\nabla \varphi)^2 \; , \label{freeghost}
\ee
which is simply that of a massless scalar field but with the ``wrong" overall sign. An overall sign does not affect the classical equations of motion of course. In fact, the theory can be quantized as well. No instability of the vacuum arises provided there is no coupling to other ``normal" particles with positive energy. As this semi-trivial counter-example indicates, it is possible to violate the NEC without being in conflict with QFT. The validity of the NEC has consequently been called into question \cite{Barcelo:2002,Rubakov:2014}; indeed, we now know of linearly stable, non-trivial examples of interacting NEC-violating theories such as theories of ghost condensates \cite{ArkaniHamed:2003} or covariant and conformal galileons \cite{Deffayet:2009wt,Nicolis:2009}. There are even theories admitting a Poincar\'e-invariant vacuum, with a Lorentz-invariant S-matrix that satisfies the dispersion relations that arise from analyticity constraints, and in which perturbations of the vacuum propagate subluminally \cite{Easson:2013bda,dbigenesis}. Thus, while all these theories violate (\ref{TNEC}), they seem perfectly consistent with quantum field theory, at least as non-renormalizable effective field theories.

The possibility of having NEC-violating matter has led to many scenarios in which such matter is minimally coupled by hand to Einstein gravity. Of course, problems arise even with (\ref{freeghost}) since gravity can mediate interactions with ``normal" particles. However, ghost instabilities can be mild \cite{Emparan:2005gg,Garriga:2012pk}. If the gravitational coupling of NEC-violating theories is valid, then the central assumption in the aforementioned gravitational no-go theorems is lifted, so that everything from bouncing cosmologies \cite{galileonbounce,ghostbounce,Easson:2011zy} to traversable wormholes becomes permissible.

The purpose of this paper is to argue that the naive minimal-coupling of NEC-violating matter to Einstein gravity is quite likely inconsistent. It appears to be in contradiction both with black hole thermodynamics and with string theory. Instead, we suggest that NEC-violating  matter could potentially be consistently coupled to gravity either if it is non-minimally coupled or if the gravitational action includes higher-curvature terms; we provide an example of each. In both cases, however, the advantage of NEC-violating matter in bypassing gravitational no-go theorems is lost.

\section{Gravitational Problems of NEC-Violating Theories}

We have seen that it is not manifestly the case that the null energy condition, (\ref{TNEC}), follows from the tenets of quantum field theory. However, when we now minimally couple matter  to Einstein gravity, Einstein's equations imply that
\be
R_{\mn} v^\mu v^\nu \geq 0  \Leftrightarrow T_{\mn} v^\mu v^\nu \geq 0 \; . \label{Rnec-Tnec}
\ee
That is, if the null energy condition, (\ref{TNEC}), holds for matter, then 
\be
R_{\mn} v^\mu v^\nu \geq 0 \; , \label{RNec}
\ee
known as the null or Ricci convergence condition, holds for geometry (and vice versa). Of course, within the framework of general relativity, there is no compelling reason why (\ref{RNec}), should hold. Thus it appears that we can derive neither the left-hand side of (\ref{Rnec-Tnec}) from general relativity, nor the right-hand side from quantum field theory. There are, however, at least two other possible sources of conditions which are neither quantum field theory nor general relativity: thermodynamics and string theory. Here we will show, by demanding the second law of thermodynamics for black holes and the consistency of worldsheet string theory, that both of these require the Ricci convergence condition to hold.

The significance of the Ricci convergence condition lies in the null version of Raychaudhuri's equation, which determines the focussing of light rays. Raychaudhuri's equation states that a null geodesic congruence with affine parameter $\lambda$, expansion parameter $\theta$, and shear tensor $\sigma$ satisfies
\be
\frac{d \theta}{d \lambda} = - \frac{1}{2} \theta^2 - \sigma^2 - R_{\mn} v^\mu v^\nu \; , \label{raych}
\ee
where we have ignored a vorticity term for simplicity. Thus if the Ricci convergence condition, (\ref{RNec}) holds, then every term on the right in Raychaudhuri's equation is negative. This is the key requirement in proving a number of gravitational theorems. In particular, it is used in Hawking's proof \cite{Hawking:1976} of the area law that says that the event horizons of classical black holes can never decrease. For when $\frac{d \theta}{d \lambda}$ is non-positive, an examination of (\ref{raych}) shows that if $\theta$ is ever negative it will become infinitely negative in finite affine time. Applied to a congruence of null generators of the event horizon, this would indicate the presence of a horizon caustic, i.e. a naked singularity on the horizon of the black hole. Cosmic censorship then dictates that $\theta$ must be positive, meaning that classical black holes cannot shrink:
\be
T_{\mn} v^\mu v^\nu \geq 0 \Rightarrow R_{\mn} v^\mu v^\nu \geq 0 \Rightarrow \frac{d \theta}{d \lambda} \leq 0 \Rightarrow \theta \geq 0 \Rightarrow \frac{dA}{d \lambda} \geq 0 \Rightarrow \frac{dS}{d \lambda} \geq 0 \; . \label{2ndlaw}
\ee
Evidently a violation of the null energy condition would break this chain of logic and could potentially allow black hole event horizons to shrink even classically. It would be nice if we could reverse the arrows to conclude that the Ricci convergence condition or the NEC are logical consequences of the second law of thermodynamics. However, it is apparent that this does not quite follow. For one thing, a spacetime that contains no event horizons would be unconstrained. Indeed, the second law can, at best, constrain the energy-momentum tensor in the vicinity of a horizon; spacetimes for which the Ricci convergence condition was violated far away from any black holes would not cause any conflict with the second law. Another difficulty is that an increase in the total area, $dA/d \lambda \geq 0$ does not imply that every area element does not shrink, $\theta \geq 0$. By contrast, the NEC and the Ricci convergence condition are local requirements. Therefore, in order to rigorously show that the NEC holds everywhere in spacetime, one would need a local version of (\ref{2ndlaw}); it would be very interesting to see whether (\ref{RNec}) could be derived from applying the second law to local Rindler horizons, perhaps similar to the way Einstein's equations themselves come out of assuming thermodynamics of local Rindler horizons \cite{Eeqnofstate,beyondE}. At present, though black holes provide strong evidence in support of the NEC, the reasoning falls short of being a proof. 

A more robust obstruction to coupling NEC-violating theories to Einstein gravity comes from perturbative string theory. (There have also been proposals, using AdS/CFT, that the null energy condition in a bulk spacetime map to a kind of c-theorem in the dual conformal field theory \cite{c-thm}. However, the validity of a c-theorem in dimensions greater than two has not been independently established, so the NEC is used to prove a c-theorem rather than vice versa.) Here we will review a direct derivation of the NEC \cite{NECderivation} using perturbative string theory. Worldsheet string theory is described by a two-dimensional nonlinear sigma model in which $D$ scalar fields, $X^\mu (\sigma, \tau)$ -- we focus here on bosonic string theory -- are minimally coupled to two-dimensional Einstein gravity on the worldsheet. For a string propagating in an arbitrary curved spacetime, the Polyakov action is
\be
S_P[X^\mu,h_{ab}] =  - \frac{1}{4 \pi \a'} \int d^2 \sigma \sqrt{-h} \left [\a' \Phi(X) R_h  +  
h^{ab} \pl_a X^\mu \pl_b X^\nu g_{\mn}(X) \right ]  \; . \label{curved-action}
 \ee
Here $h_{ab}$ is the metric on the two-dimensional worldsheet and $R_h$ is its Ricci scalar. The background fields are $g_{\mn}(X)$, the metric of $D$-dimensional spacetime, and $\Phi(X(\tau, \sigma))$, the dilaton; we neglect the anti-symmetric Kalb-Ramond field, $B_{\mn}$, for simplicity.

We now perform a background field expansion $X^\mu(\tau, \sigma) = X^\mu_0 (\tau, \sigma) + Y^\mu (\tau, \sigma)$ where $X^\mu_0 (\tau, \sigma)$ is some solution of the classical equation of motion for $X^\mu$. Then, for every value of $(\tau, \sigma)$, we can use standard field redefinitions \cite{CallanThorlacius,GSW1} to expand the metric in Riemann normal coordinates about the spacetime point $X^\mu_0 (\tau, \sigma)$:
\be
g_{\mn} (X) = \eta_{\mn} - \frac{1}{3} R_{\mu \a \nu \beta} (X_0) Y^\a Y^\beta + ... \; . \label{RNC}
\ee
Contracted with $\partial_a X^\mu \partial^a X^\nu$, the second and higher terms introduce quartic and higher terms in the Lagrangian turning (\ref{curved-action}) into an interacting theory. The resultant divergences can be cancelled by adding suitable counter-terms to the original Lagrangian. Integrating out $Y$, the one-loop effective action is
\cite{CallanThorlacius,GSW1}
\be
S[X_0^\mu,h_{ab}] = - \frac{1}{4 \pi \a'} \int d^2 \sigma \sqrt{-h} \left [  \a' C_\epsilon \Phi R_h 
+ h^{ab} \pl_a X_0^\mu \pl_b X_0^\nu (\eta_{\mn} + C_\epsilon \a' R_{\mn}) \right ] \; . \label{eff-action}
\ee
Here $C_\epsilon$ is the divergent coefficient of the counter-terms. 

Consider now the equation of motion for the worldsheet metric, $h_{ab}$. Employing light-cone coordinates on the worldsheet,
\be
\sigma^{\pm} \equiv \tau \pm \sigma \; ,
\ee
the equation of motion reads
\be
0 =  \pl_{\pm} X_0^\mu  \pl_{\pm} X_0^\nu \left ( \eta_{\mu \nu} + C_\epsilon \alpha' ( R_{\mu \nu} + 2 \nabla_\mu \nabla_\nu \Phi ) \right ) \; . 
\ee
These are the Virasoro constraints. Since $C_\epsilon$ is divergent and cut-off dependent, the terms that do and do not involve $C_\epsilon$ must vanish separately. For the $C_\epsilon$-independent term, we have
\be
0 =  \pl_+ X^\mu \pl_+ X^\nu \eta_{\mn} \; ,
\ee
with a similar equation with $+$ replaced by $-$. Defining a vector field $v_+^\mu = \pl_+ X^\mu(\sigma, \tau)$, we find that
\be
\eta_{\mn} v_+^\mu v_+^\nu = 0 \; , \label{nullv}
\ee
which is to say that $v_+^\mu$ is a null vector field. Thus, worldsheet string theory naturally singles out spacetime null vectors. To derive the Ricci convergence condition, consider then an arbitrary null vector $v^\mu$ in the tangent plane of some arbitrary point in an arbitrary spacetime. Let there be a test string passing through the given point with $\pl_+ X^\mu$ equal to $v^\mu$ at the point. The terms that depend on $C_\epsilon$ then read
\be
v_+^\mu v_+^\nu (R_{\mn} + 2 \nabla_\mu \nabla_\nu \Phi  ) = 0 \; . \label{alphavirasoro}
\ee
This is very nearly the Ricci convergence condition, (\ref{RNec}), except for two differences: it is an equality, rather than an inequality, and there is an extra, unwanted term involving the dilaton.

However, the metric, $g_{\mn}$, that appears in the worldsheet action is the string-frame metric. We can transform to Einstein frame by defining:
\be
g_{\mn} = e^{\frac{4 \Phi}{D-2}} g_{\mn}^E \; . \label{weyl}
\ee
Then we find that
\be
R^E_{\mn} v_+^\mu v_+^\nu  = + \frac{4}{D-2} (v_+^\mu \nabla^E_\mu \Phi )^2 \; .
\ee
The right-hand side is manifestly non-negative. Hence we have
\be
R^E_{\mn} v_+^\mu v_+^\nu  \geq 0 \; .
\ee
This establishes the Ricci convergence condition \cite{NECderivation}, which is equivalent to the null energy condition when Einstein's equations hold. It is not difficult to show that this result holds in all dimensions \cite{NECderivation}.

\section{How To Couple NEC-Violating Theories of Matter to Gravity}
\label{nec-violation}

We have seen that black hole thermodynamics provides evidence for, and string theory requires, background spacetimes that satisfy the Ricci convergence condition. It may appear then that effective theories of matter that violate the null energy condition (in the form $T_{\mn} k^\mu k^\nu \geq 0$) can immediately be ruled out. However, there remains a loophole. When we write down a theory of matter that violates the null energy condition, we mean violation of (\ref{TNEC}). What appears in string theory and the black hole area law, however, is (\ref{RNec}). These are equivalent only if the gravitational equations are Einstein's equations (under which we include a possible cosmological constant). Thus NEC-violating matter might be allowed if it does not couple to gravity in the conventional way. For example, the NEC-violating matter might be non-minimally coupled. Or, there could be additional higher-curvature terms in the gravitational action. To be clear, NEC-violating theories are typically effective theories which have usually been coupled by hand to Einstein gravity. We propose that these theories instead be gravitationally coupled, again by hand, either non-minimally or to higher-curvature terms. This would at least retain the possibility of being consistent with black holes and string theory.

Let us see how these modifications evade the difficulties with both black holes and string theory. First consider non-minimal coupling. This could be for example some function of a scalar field multiplying the Ricci scalar, or a kinetic coupling like $R^{ab} \partial_a \varphi \partial_b \varphi$, or something else. To be specific, we will consider a Jordan frame, in which a scalar field is coupled to the Ricci scalar via a term $\Omega^{-2}(x) R$. In general, when matter is non-minimally coupled, the distinction between matter and gravity is ambiguous: the action does not separate into distinct matter and gravity parts. Correspondingly, the energy-momentum tensor is ambiguous. One possible way to define it is to equate it to the Einstein tensor:
\be
G^J_{\mn} \equiv T_{\mn}^J \; .
\ee
This ensures that $T_{\mn}^J$ is covariantly conserved, but the right-hand side will now generically include terms that depend on spacetime curvature. Now, in Jordan frame, there is still of course a condition that comes from the second law, but it does not take the usual form of the NEC. By comparison with Einstein frame, it can be shown \cite{jones} that, to guarantee the second law, the needed condition is
\begin{eqnarray}
v^\mu v^\nu \left ( T^J_{\mn} + \nabla_\mu \nabla_\nu \ln \Omega \right ) \geq 0 \; , \label{NECJ}
\end{eqnarray}
for any null vector, $v^\mu$. Thus, it can be the case that $T_{\mn}^J$ fails to satisfy the NEC in its usual form, (\ref{TNEC}), without being inconsistent with black hole thermodynamics so long as (\ref{NECJ}) is obeyed. Non-minimal coupling also eliminates the obstruction from string theory. It is easiest to see this by going back to string frame (in which the dilaton is non-minimally coupled): we then recover (\ref{alphavirasoro}), which does not take the form of the Ricci convergence condition.

Next consider how higher-curvature terms in the gravitational action remove the NEC as a condition. A proof of the second law does not presently exist for black holes of higher-curvature gravity. Nevertheless, it is clear that the second law no longer calls for the Ricci convergence condition. This is because, since the Raychaudhuri equation is valid for all theories of gravity, if the Ricci convergence condition were obeyed, it would again imply that $d A / d\lambda \geq 0$. However, in generalized theories of gravity, this is not the correct condition; in such theories, the black hole entropy is not proportional to the area but is given instead by the Wald entropy \cite{wald},
\be
S = \frac{1}{8 \kappa} \int dS_{\mn} J^{\mn} \; ,
\ee 
where $\kappa$ is the surface gravity and $J_{\mn}$ is a Noether current. Demanding that the Wald entropy increase with time will again imply some condition on spacetime geometry, but it will not be the Ricci convergence condition. (This is similar to the situation with non-minimal coupling; indeed, for $f(R)$ gravity, the theory can be regarded either as a higher-curvature gravity theory or as a non-minimally coupled scalar-tensor theory.) Furthermore, the presence of additional terms in the gravitational equations will generically break the equivalence (\ref{Rnec-Tnec}). The net result is that the validity of the null energy condition does not follow (even with all the caveats of the previous section) from the generalized second law in higher-curvature theories of gravity.

In the context of string theory, the inclusion of higher-curvature terms is tantamount to considering higher order terms in the $\a'$ expansion. These will generically add terms proportional to quadratic and higher powers of $\a'$ to the Virasoro constraint, (\ref{alphavirasoro}). This will again lead to some condition, but it will not be the Ricci convergence condition. A specific example of a NEC-violating theory that is not in manifest conflict with string theory because of higher-curvature gravity is conformal galileon theory. This is because conformal galileons have been shown to be derivable by Kaluza-Klein compactification of Lovelock gravity \cite{galileonsfromGB}, of which the Einstein and Gauss-Bonnet pieces are contained in the low-energy effective action of heterotic string theory. However, since the conformal galileons are the Kaluza-Klein scalars under dimensional reduction of Einstein-Gauss-Bonnet gravity they are not coupled to Einstein gravity alone but to a higher-curvature theory, namely Einstein-Gauss-Bonnet gravity \cite{galileonsandgravity}. Correspondingly, a contraction of the equation of motion with null vectors gives $R_{\mn} v^\mu v^\nu + c H_{\mn} v^\mu v^\nu  = T_{\mn} v^\mu v^\nu$, where $H_{\mn}$ is the Gauss-Bonnet contribution to the gravitational equation of motion. The presence of $H_{\mn}$ severs the connection between the NEC and the Ricci convergence condition. While this permits galileons to be coupled to gravity, it is important to note that, in terms of evading gravitational no-go theorems such as those that prohibit wormholes or bouncing cosmologies, the benefits of violating the NEC are lost because of the higher-curvature terms.

In summary, there are two independent lines of evidence, coming from black holes and worldsheet string theory, for the Ricci convergence condition. Here we have considered the implications for the gravitational coupling of NEC-violating matter. In the absence of a UV-complete theory that contains gravity and definitively reduces to a NEC-violating effective theory \cite{Adams:2006sv}, the precise relation between NEC-violating matter and gravity is unclear. In much of the literature,  NEC-violating matter has simply been coupled by hand to Einstein gravity. We have argued here that this is incompatible with string theory and black hole thermodynamics, as a corollary to the necessity of the Ricci convergence condition. But we have also identified methods, using non-minimal couplings and higher-curvature gravity, for coupling NEC-violating matter to gravity in a way that is not manifestly inconsistent with string theory and black hole thermodynamics.

\bigskip
\noindent
{\bf Acknowledgments}

\noindent
M.P. is supported in part by DOE grant DE-FG02-09ER41624. J.P.v.d.S. acknowledges the Delta ITP consortium, a program of the Netherlands Organization for Scientific Research (NWO), which is funded by the Dutch Ministry of Education, Culture and Science (OCW).

\end{document}